\documentclass[12pt,notitlepage]{article}
\usepackage{amssymb}
\usepackage{amsmath}
\usepackage{pdfpages}
\usepackage{graphicx}
\usepackage{epstopdf}
\usepackage{pdflscape}
\usepackage{tabularx}
\usepackage{longtable}
\usepackage{hyperref}
\usepackage{breakurl}
\usepackage{amsthm}
\usepackage[margin=10pt,font=small,labelfont=bf, labelsep=endash]{caption}
\usepackage{float}
\usepackage{rotating, graphicx}


\begin{document}

\title{Pandemic, Shutdown and Consumer Spending: Lessons from Scandinavian Policy Responses to COVID-19\thanks{%
Asger Lau Andersen: asger.lau.andersen@econ.ku.dk; Emil Toft Hansen: eth@econ.ku.dk; Niels Johannesen:
niels.johannesen@econ.ku.dk; Adam Sheridan: adam.sheridan@econ.ku.dk. We are extremely grateful to key employees at Danske Bank for their help. All individual data used in this analysis has been anonymized and no single customer can be traced in the data. All data processing has been conducted by authorized Danske Bank personnel, following the bank's strict data privacy guidelines.} }

\author{Asger Lau \textsc{Andersen} (University of Copenhagen and CEBI) \\
Emil Toft \textsc{Hansen} (University of Copenhagen and CEBI) \\
Niels \textsc{Johannesen} (University of Copenhagen and CEBI) \\
Adam \textsc{Sheridan} (University of Copenhagen and CEBI)\\
\ }
\date{\today \\
}
\maketitle

\begin{abstract}
\noindent


\noindent This paper uses transaction data from a large bank in Scandinavia to estimate the effect of social distancing laws on consumer spending in the COVID-19 pandemic. The analysis exploits a natural experiment to disentangle the effects of the virus and the laws aiming to contain it: Denmark and Sweden were similarly exposed to the pandemic but only Denmark imposed significant restrictions on social and economic activities. We estimate that aggregate spending dropped by around 25 percent in Sweden and, as a result of the shutdown, by 4 additional percentage points in Denmark. This implies that most of the economic contraction is caused by the virus itself and occurs regardless of whether governments mandate social distancing or not. The age gradient in the estimates suggest that social distancing reinforces the virus-induced drop in spending for individuals with low health risk but attenuates it for individuals with high health risk by lowering the overall prevalence of the virus in the society.

\end{abstract}

\thispagestyle{empty}\baselineskip1.4\baselineskip \newpage{}

\pagestyle{plain}\pagenumbering{arabic}

\clearpage

\newpage


\section{Introduction}
\label{sec:intro}

One of the key policy choices facing governments in the COVID-19 pandemic is whether to shut down economic activity in order to slow the spread of the disease. Many types of consumption take place in settings with a high density of people (e.g., restaurants, public transit) or involve direct physical contact (e.g., hair dressers, dentists) and contribute to the disease spreading. Other types of consumer spending (e.g., retail shopping) involve proximity to shop assistants and other consumers and therefore also pose risks. Most governments have implemented social distancing laws that restrict some of these economic activities. However, the severity of these restrictions vary considerably across countries and many governments are considering day-by-day whether to loosen, maintain, or tighten restrictions in response to new data on COVID-19 cases and mortality.

The policy choice facing governments is often portrayed as a simple trade-off between saving lives and saving the economy. By this logic, more severe restrictions help contain the virus and reduce the ultimate death toll but cause more economic pain as more households cut spending, more businesses go bankrupt and more workers lose their jobs.\footnote{For instance, Gourinchas (2020) writes ``flattening the infection curve inevitably steepens the macroeconomic recession curve."} However, this view does not account for the possibility that the underlying health risks that the restrictions are designed to curtail may inflict significant harm on the economy in their own right: individuals may themselves choose to restrain economic activity based on their personal health risk, even when this is not mandated by law. This points to an indirect mechanism whereby shutting down sectors with high social proximity may \textit{improve} aggregate economic outcomes: it constrains the choices of low-risk individuals (e.g., the young) who would otherwise contribute most to the spreading of the disease and thus induces high-risk individuals (e.g., the old) to choose a higher level of economic activity.

In this paper, we study empirically how social distancing laws affect aggregate consumer spending in a pandemic: how much spending is lost when sectors with high social proximity are shut down compared to a situation where these sectors remain open and the disease is contained less effectively? We also investigate an important mechanism underlying the effect of shutdowns by estimating how the effect varies across groups with different health risks: do social distancing laws cause particularly large drops in spending for low-risk individuals while allowing for greater economic activity for high-risk individuals as a result of the containment of the virus?

Our empirical design draws on a salient natural experiment in Scandinavia. The two neighboring countries, Denmark and Sweden, were exposed to the COVID-19 pandemic in a very similar way. In both countries, the epidemic started on 26-27 February 2020 and mortality rates increased at a similar pace over the following weeks.\footnote{Sweden had a single confirmed case earlier on 31 January 2020, which was fully isolated.} The policy response, however, differed significantly in the two countries. Starting on 11 March 2020, Denmark took measures to limit social interaction in order to contain the virus: congregations of more than 10 people were prohibited; schools, universities and other non-essential parts of the public sector were shut down; borders were closed for foreign nationals; and the population was encouraged to stay at home and minimize social contact. The measures had direct implications for many types of economic activity involving high social proximity: \textit{restaurants} were not allowed to seat customers; the \textit{entertainment} industry including nightclubs, cinemas and concert venues was shut down; \textit{personal services} such as hair dressers and dentists were closed; \textit{retail} was restricted to high-street shops with malls being shut down entirely; and \textit{public transport} was limited. Compared to Denmark and almost every other Western country, Sweden took a much lighter touch approach: the population were encouraged to stay home if feeling unwell and to limit social interaction if possible; large-scale events were prohibited, and restaurants and bars restricted to table service only, but private businesses were generally allowed to operate freely. Consistent with the notion that restrictive measures help reign in the pandemic, mortality rates in Denmark and Sweden diverged sharply around two weeks after the Danish shutdown.

Because Denmark and Sweden were exposed similarly to the epidemic, but only the Danish government restricted economic activity, a comparison of spending dynamics in the two countries allows us to isolate the effect of the restrictions. Importantly, the comparison captures both the \textit{direct} effects of the restrictions through the reduced availability of goods and services and the \textit{indirect} effects of the restrictions through the reduced spreading of the virus. By isolating the differential effect of government policy, our analysis is fundamentally different from recent papers estimating the total change in consumer spending through the COVID-19 pandemic in the United States (Baker et al., 2020), China (Chen et al., 2020), Spain (Carvalho et al., 2020), France (Bounie et al., 2020), the United Kingdom (Hoke et al., 2020) and Denmark (Andersen et al., 2020a).

Our analysis is based on transaction data for about 860,000 individuals in Denmark and Sweden who are active customers at Danske Bank, the second-largest bank in Scandinavia with a customer base that is broadly representative of the general population. For each customer, we measure spending by cards, cash withdrawals, mobile wallets and settlements of online invoices until 5 April 2020. This allows us to construct a comprehensive customer-level measure of spending at the daily frequency, which accommodates substitution across modes of payment.\footnote{Most of the existing studies of consumer spending through the COVID-19 crisis are based on card payments alone, which is problematic if there are large shifts in modes of payment. For instance, we document a significant drop in the use of cash suggesting that consumers responded to the health risk inherent to touching coins and bills that frequently change hands.} Exploiting a standardized classification of merchants, we break down card spending by expenditure categories and add them up to categories of spending that vary in their degree of social proximity, such as \textit{social spending} in restaurants, cinemas and concert venues, \textit{personal care} spending at hairdressers, beauticians and dentists and \textit{retail shopping} on the high street and in malls. Finally, the dataset allows us to distinguish between different demographic segments based on customer information about age and gender.

Our first result is that the Danish shutdown had a modest effect on aggregate spending. We estimate that Danish customers reduced spending by almost 29\% relative to a counterfactual trajectory without the COVID-19 pandemic. This is enormous compared to spending drops in other contexts, e.g., it is four times larger than the typical spending drop around unemployment events (Ganong and Noel, 2019; Andersen et al., 2020b), but only around half the size of the aggregate spending drop in countries more adversely affected by the pandemic, such as France (Bounie et al., 2020) and Spain (Carvalho et al., 2020). By comparison, we estimate that Swedish customers reduced spending by around 25\%.\footnote{This estimate (24.84\%) is based on weighting of the Swedish observations to match the socio-demographic (age, gender, income) composition of the Danish sample. The estimate based on the unweighted observations is 24.72\%.} The difference of just under 4 percentage points is precisely estimated and represents our preferred estimate of the causal effect of the Danish shutdown on aggregate spending. 

This finding suggests that, in the context of an escalating pandemic, shutting down sectors with high social proximity is associated with a relatively small loss of aggregate spending. Even when there are no major restrictions on economic activity, as in Sweden, a pandemic induces a sizeable contraction of spending. The additional drop in spending caused by a shutdown, as in Denmark, is relatively small. While these results confirm the popular notion that the lives saved by a shutdown come at an economic cost, the trade-off is much less stark than suggested by the large drops in economic activity around shutdowns. 

Our next results document that the small effect of the shutdown on aggregate spending conceals considerable heterogeneity across groups with different health exposure to the virus. Specifically, we find that the effect of the shutdown on total spending exhibits a significant age gradient: the effect is negative for young adults (18-29 years) and positive for the oldest group (70+ years). When breaking total spending into subcategories, we observe a clear positive age gradient across the four major spending categories involving high social proximity: retail (i.e., high-street shops and malls), social spending (i.e., food, drink and entertainment away from home), personal care (e.g., hairdressers and dentists) and public transport (i.e., trains and busses). As the risk of suffering serious health consequences of COVID-19, hospitalization and ultimately death, is strongly increasing in age, we interpret the age gradient as the manifestation of an underlying health risk gradient in the spending effects. 

These results suggest that, in a pandemic, a shutdown \textit{decreases} the total spending of individuals with low health risk. The effects are particularly pronounced for categories involving high social proximity such as personal care and social spending. While low-risk individuals significantly reduce spending that exposes them to the virus even without any government interventions, a shutdown effectively constrains their choices and makes them reduce such spending even further. 

The results also suggest that, in a pandemic, a shutdown \textit{increases} the total spending of individuals with high health risk. The overall increase reflects modest decreases in spending for categories involving high social proximity (e.g., personal care and social spending), and increases for categories involving moderate social proximity (e.g., retail shopping and public transport). In other contexts, it would be highly surprising that constraints on the availability of certain goods and services cause some individuals to choose a higher overall level of consumption. In the context of the pandemic, however, shutting down sectors with high social proximity reduces the prevalence of the virus in society at large. This implies a lower risk of contracting the virus in, for instance, public transit and retail shops, which stimulates the spending of individuals with a high underlying health risk.  

Our findings have important implications for the appropriate policy response to a pandemic by suggesting that most of the reduction in economic activity is caused by the pandemic itself and occurs regardless of whether governments shut down sectors with high social proximity or not.\footnote{This is consistent with historical evidence that public health interventions to contain flu pandemics reduced mortality (Bootsma and Ferguson, 2007) without depressing the economy (Correia et al., 2020). It is also consistent with recent evidence from U.S. states that working hours (Bartik, 2020) and consumer spending (Chetty et al., 2020) dropped significantly before state-level shutdowns.} Governments should weigh the benefits of the public health interventions in terms of reduced mortality  and serious illness (Juranek and Zoutman, 2020) against the small differential cost in terms of economic activity.

Our findings also inform the rapidly developing macroeconomic literature on pandemics. The models start from different assumptions about the nature of the shock to the economic system. Some assume that the COVID-19 pandemic is essentially a shock to the supply side with possible spill-overs to the demand side (Guerrieri et al., 2020), while others emphasize that the pandemic affects demand directly because it introduces a health cost of consumption (Eichenbaum et al, 2020). Our findings suggest that the direct demand shock is important: spending drops massively even when supply is unconstrained and the drop correlates strongly with health risk.\footnote{We cannot exclude that some of the drop in spending that we estimate in Sweden and Denmark is due to merchants closing down in response to concerns about the health risks for workers. This could be considered a supply shock.}

Finally, our analysis contributes to an emerging literature studying the effects of the social distancing laws imposed by most governments in the world in response to the COVID-19 pandemic. Most of the literature focuses on the effectiveness of these policies in limiting personal mobility (Brzezinski et al., 2020; Sears, 2020) and containing the virus (Kraemer et al., 2020). While several papers use quantitative models to evaluate social distancing policies (e.g. Bodenstein et al., 2020; Acemoglu et al., 2020), we are not aware of other causal evidence on the effect on spending or other dimensions of economic activity.

The paper proceeds in the following way. Section 2 briefly accounts for the Danish context. Section 3 describes the data sources and provides summary statistics. Section 4 develops the empirical framework. Section 5 reports the results. Section 6 concludes.


\section{Background}
\label{sec:background}

Denmark and Sweden had very similar early experiences of the COVID-19 outbreak, with cases first taking off in the last week of February 2020. In both countries early cases were concentrated among people returning from travels in the most affected areas of Europe, particularly Northern Italy, with community spread following soon after. 

Figure 1 provides evidence on the similar early experiences of COVID-19 in Sweden and Denmark. The top panel plots the evolution of weekly excess deaths in 2020 in Denmark, Sweden and, as a point of contrast, Italy. Excess deaths are measured as the percentage difference in all-cause mortality in a given week relative to average deaths in that week over the previous 5 years. Through January and February, and up until early March, excess deaths evolved very similarly in Denmark and Sweden: mortality was generally below normal, suggesting a milder flu season than average, with some hints of an increasing trend in the first week of March in both countries. In contrast, mortality was rising in Italy from mid-February and excess deaths reached almost 15\% in the first week of March. 

To provide further evidence on the similarity of early experiences, the bottom panel of Figure 1 shows the evolution of excess weekly Google searches for the terms ``cough", ``fever", and ``sick".\footnote{In Danish, Swedish, and Italian, respectively: ``cough" (``hoste",``hosta", ``tosse"), ``fever" (``feber", ``feber", ``febbre"), ``sick"  (``syg", ``sjuk", ``malato").} In each country, Google searches for these terms display a seasonal pattern, peaking in winter at the height of flu season. This seasonality presumably reflects people Googling their symptoms. Thus, excess searches for these terms, relative to the average search behaviour in weeks in previous years, suggests an increase in the incidence of illness. Moreover, the symptoms of COVID-19 -- especially a cough and a fever -- were heavily publicized in the media and by public health authorities as the outbreak took hold. Therefore, excess searches for these terms in early 2020 may reflect both signs of an outbreak and heightened awareness of COVID-19. As Figure 1 demonstrates, Google searches for these terms began rising in Sweden and Denmark at precisely the same time in mid-February. By contrast, the series for Italy shows that Google searches for these terms were already above normal in January, highlighting how these data capture the differential timing of outbreaks across countries. 

Despite these similar early experiences, Denmark and Sweden adopted very different policy responses to the outbreak. On 11 March, the prime minister of Denmark announced a national lockdown in a televised speech: schools, universities and other non-essential parts of the public sector were shut down; borders were closed for foreign nationals, effectively ending international travel; employers were urged to allow their employees to work from home, and people were urged to stay at home and minimize social contact. On 18 March, the government announced further restrictions banning congregations of more than 10 individuals, shutting down shopping malls, closing establishments of high physical proximity, such as nightclubs and hairdressers, limiting healthcare practices such as dentists to emergency treatment only, and restricting restaurants, cafes and bars to take-away service only. The timing and severity of the measures were generally comparable to most of Northern Europe (such as Germany and the Netherlands), but less restrictive than in Southern Europe where the outbreak was more severe (such as Italy and Spain).

In contrast to Denmark, and very different to almost all other Western countries, the Swedish government opted for a lighter-touch approach to manage the outbreak, with most measures being voluntary and coming relatively late in the outbreak.\footnote{Swedish Health Minister Lena Hallengren was quoted as saying that their ``strategy has always been to introduce the measures at the time when they were necessary, at the point in the spread of infection when we have noticed they are needed - perhaps a bit later than in other countries - but that's been our aim." (The Local, 2020)} For example, the government first advised people to stay at home if feeling sick on 10 March, the day before Denmark went on lockdown. On 16 March, the government issued their first recommendations to employers to allow their employees to work from home, and for people aged over 70 to limit close contact with others. On 17 March, the government recommended that universities and senior high schools (for children aged over 16) move to distance learning. Moreover, the government issued their first recommendation for people to practice social distancing by avoiding meeting friends and relatives in person, if possible, on 24 March. These key examples demonstrate the focus of the government on voluntary measures to tackle the outbreak, but they did also introduce a number of legal restrictions as things progressed. Meetings of more than 500 people were outlawed on 11 March, falling to 50 people on 27 March. On 24 March the government restricted all bars, restaurants and cafes to table service only. Finally, on 1 April the government announced a set of guidelines including that individuals should keep distance from others in public and that high-risk groups, including over 70s, should avoid going shopping and meeting people. The government did not impose any sanctions for not following these guidelines.   

Figure 1 also demonstrates the consequences of these different policy responses for public health. As shown in the top panel, two weeks after the Danish shutdown, marked by the dashed line on 11 March, excess deaths began to diverge, with a sharp and continuous increase to over 40\% in Sweden (coming close to levels in Italy) but a levelling off at around 5\% in Denmark. Moreover, the bottom panel shows that excess Google searches for symptoms began to fall quite quickly in Denmark following the shutdown, but have remained elevated in Sweden. Figure 1 ends on 12 April, the last week of data on mortality that is available for Italy at the time of writing. But for Denmark and Sweden, the most recent data suggests that excess weekly mortality has remained at around 40\% in Sweden and has fallen to around 2\% in Denmark.   

This evidence suggests that Sweden and Denmark would have continued to experience similar COVID-19 outbreaks, had they followed the same policy. The decision by the Swedish government to remain open was not motivated by differences in the severity of the COVID-19 outbreak and, as will be seen in our empirical results, by differences in economic performance. One factor that may explain this exogenous difference in policy is a historical feature of the Swedish constitution that is not present in Denmark: the Swedish constitution does not allow the government to call a state of emergency in peacetime, making it difficult - and historically unprecedented - to quickly pass laws affecting individual liberties (Karlson, Stern and Klein, 2020).


\section{Data}
\label{sec:data}

We study the effect of COVID-19 shutdown policies on consumer spending by using bank account data from Danske Bank. The dataset for this paper is drawn from the 5 million adult individuals across the Nordic countries with an account at the bank. This data has three advantages over existing data sources. First, it includes information on the spending and key demographics of a large sample of individuals. Second, it spans two countries, Denmark and Sweden, which are highly comparable in terms of population, institutions and initial exposure to the COVID-19 outbreak but which differ dramatically in terms of government policy response to the outbreak. Third, for many years Denmark and Sweden have had the highest card payments per capita and the lowest cash payments per capita of almost any other countries (Danmarks Nationalbank, 2016), meaning that our measures of spending are precise and comprehensive. 

Our key outcome variable is daily total consumer spending. We measure total spending as the sum of credit and debit card transactions, mobile wallet payments, cash withdrawals, and electronic invoice payments associated with online shopping.\footnote{From card payments, we exclude transactions that are identified as tax repayments or financial services, such as money transfers to other persons or debt repayments.} Electronic invoices are a common payment option at online merchants in Sweden. Since our focus is on high-frequency spending dynamics, we exclude other bill payments from our spending measure. These other bill payments include things like direct debits for utilities that can be adjusted less quickly and can have a large disparity between timing of payment and timing of consumption.

Beyond total spending, we are interested in studying how the Danish shutdown affects spending behaviour in venues that vary in their level of (actual or perceived) social and physical distancing and, consequently, in the extent of government intervention on their activity. To this aim, we focus on offline card transactions and create four spending categories of interest.\footnote{Every card and mobile wallet transaction is associated with a Merchant Category Code (MCC), an international standard for categorising merchants according to the goods or services they provide. We use these MCCs to construct our spending measures.} First, \textit{social spending} includes payments at venues of particularly high social proximity, either because they are venues where a lot of socialising takes place or because they involve prolonged indoor exposure to others. Examples of social spending include payments on food and drink away from home, cinemas, museums and galleries, theatres, and sports clubs. Second, we measure spending on \textit{personal care services}, a type of spending that is characterised by one-on-one physical contact and includes services such as hairdressers and beauticians, and spending on medical services, such as dentists and opticians. Third, we measure spending on \textit{public transport} as purchases of bus, train, and metro transportation; all venues of potentially high social proximity. Finally, we create a category including all \textit{high street and mall} shopping. Examples of spending in this category are clothes and electronics stores, and professional services such as auto repair and dry cleaning. This category of spending is likely to involve some proximity to merchants and other individuals and, especially for the young, may also be a social activity. We provide more detail on the coding of each spending measure in Table A1 in the Appendix.

These spending data have significant advantages over survey-based consumption measures but still present some challenges for our analysis that we deal with through careful selection of the analysis sample. First, we only observe data on spending from one bank, meaning that we may miss some spending of customers in the full sample who have multiple bank relationships. In order to minimise this problem, we limit our attention to a sample of individuals who have been customers of the bank in every month from January 2018 to December 2019, have made at least one card payment in each and every month of that two year period and, where information is available, have declared their Danske Bank account to the government as their primary account for tax purposes. We only impose the minimum monthly spending requirement until December 2019 as we want to allow for individual spending to fall to zero in response to the crisis. Second, individual spending is often partly on behalf of other household members, for example, a person might purchase flights from their personal account for themselves and their partner. In order to account for household structure, we use the bank's household identifiers -- constructed from information on joint accounts and co-residence -- in order to split all expenditures of couples on their personal and joint accounts across each spouse equally.\footnote{For instance, when one member of a household spends \$50 on groceries, we consider that each spouse has spent \$25.} Finally, in order to focus on comparable samples that vary only by government policy response to the pandemic, we limit our sample to individuals resident in either Denmark or Sweden as of December 2019. With these restrictions, our sample consists of around 860,000 people, with 760,000 people in Denmark and 100,000 people in Sweden.

We link the spending data with key demographics for each customer, sourced from the bank's customer records. These include age, gender, and permanent income. Permanent income is measured as average monthly total spending over a long period (two years). This measure is designed to capture access to economic resources for individuals at different life stages more accurately than income measured at a point-in-time.\footnote{For the Danish subsample, we also measure total disposable income from account inflows in 2019. Danske Bank categorizes account inflows and we take the sum of inflows labelled as salary, government transfers, capital income, and pension. We do not observe a similar classification for the Swedish subsample but we are able to impute their disposable income from their spending based on the average propensity to consume in the Danish sample.} The demographics serve three roles. First, they allow us to assess the representativness of our sample. Second, since they are so important in explaining cross-sectional variation in spending, we use them to assess the comparability of our sample across countries and re-weight our estimates to adjust for any imbalances, ensuring that such imbalances are not driving our results. Finally, age is a key factor in individual health risk and we explore how the effects of the shutdown vary across individuals of different ages and hence different personal exposure to the COVID-19. 

Table 1 reports summary statistics for our estimation sample and the populations they are drawn from. Column 1 reports statistics for the Danish subsample and Column 2 provides a comparison to the full adult population in Denmark, obtained from government registers. Our sample of 760,000 individuals is largely representative of the adult population of 4,600,000 in terms of gender, age, and income. This reflects that Danske Bank is the largest bank in Denmark, catering to all types of customers and with a significant presence in all parts of the country. Columns 3 and 4 report summary statistics for our Swedish subsample and a comparison to the full adult population in Sweden, again sourced from government registers. The Swedish subsample represents a significantly smaller share of the full population than the Danish subsample. This reflects the smaller relative market share of Danske Bank in Sweden. The Swedish subsample overrepresents those aged 30 to 59. The female share is consistent with that of the Swedish population and average income is slightly higher. Overall, the Swedish subsample is broadly similar to the Danish subsample and, crucially, contains large numbers of individuals across all ranges of demographics, allowing us to effectively control for possible confounding differences in response to the crisis.


\section{Empirical strategy}
\label{sec:empirical strategy}

The main aim of the empirical analysis is to measure the change in consumer spending induced by the COVID-19 crisis in Denmark and Sweden and to isolate the impact of the shutdown in Denmark separately from the impact of the pandemic itself. 

Our unit of observation is individual-by-day, allowing us to capture sharp changes in spending as a result of the pandemic and the shutdown. The high frequency creates challenges, however, due to the strong cyclicality of spending over the week, the month, and the year. As in Andersen et al. (2020a), we address the cyclicality by comparing consumer spending on each day in 2020 to consumer spending on a reference day 364 days earlier. The reference day is always the same weekday and almost exactly the same place in the monthly and annual spending cycle. For example, we compare 17 January 2020 (a Friday) to the reference day 18 January 2019 (also a Friday). This method does not account for the fact that increases in spending around paydays may fall on different weekdays in different years, but this will not affect our key estimates as explained below.\footnote{There is no uniform payday in Denmark and Sweden but salary payouts typically occur over a few days toward the end of the month.} 

For each individual in our sample and for each day of our window of analysis, we compute the difference between spending on the day itself and spending on the reference day the year before. Scaling with the average daily spending per person in the individual's country over a long period before the window of analysis, we obtain an individual-specific measure of excess spending on a given day expressed as a fraction of the normal level of spending per person in the country:

\[
excess\:spending_{ict}=\frac{spending_{ict}-spending_{ict-364}}{average\:spending_{c}}
\]

where $spending_{itc}$ is spending on day $t$ for individual $i$ living in country $c$, and $average\:spending_{c}$ is average daily spending per person in that country taken over all days in 2019.

We measure the individual-level spending change over the COVID-19 crisis as the difference between average excess spending in the post-shutdown period, 11 March -- 5 April, and average excess spending in the early pre-shutdown period, 2 January -- 15 February:

\[
\Delta spending_{ic}=\underbrace{E_{t}[excess\ spending_{ict}|t\in post]}_{\substack{\text{average \ excess \ spending}\\
\text{post-shutdown}
}
}\ -\ \underbrace{E_{t}[excess\ spending_{ict}|t\in pre]}_{\substack{\text{average \ excess \ spending}\\
\text{pre-shutdown}
}
}
\]

Equipped with this machinery, we measure the effect of the COVID-19 crisis in country $c$ by averaging over individuals in our sample from that country:

\[
\Delta spending_{c}=E_{i}[\Delta spending_{ic}]
\]

\noindent Since excess spending is measured relative to average daily spending across all individuals, $\Delta spending_{c}$ measures the impact on \textit{aggregate} spending relative to the 2019 baseline. With this measure, we effectively use excess spending in the pre-shutdown period as a counterfactual for excess spending in the post-shutdown period in each country. Simply put, we assume that year-over-year spending growth between 2019 and 2020 would have been the same after 11 March as before absent the epidemic and the shutdown. However, we exclude 16 February - 11 March from the pre-shutdown period as early restrictions (e.g. on air travel to Asia) and anticipation of the broader crisis may have affected spending prior to the shutdown. While payday spending creates spikes in excess spending on individual days -- positive when we compare a pay day to a normal day and negative when we do the opposite -- they do not affect $\Delta spending_{c}$ because both its terms average over the same number of positive and negative pay day spikes.

While $\Delta spending_{c}$ remains our summary measure of the spending effect in country $c$, we also show plots that compare spending on each day in the window of analysis to spending on the reference day the year before. These plots allow us to visually assess whether consumer spending behaved similarly in the pre-shutdown period as on the same days the year before (except for a level shift). This is key to assessing the credibility of our identifying assumption that consumer spending would have behaved similarly in the post-shutdown period as on the same days the year before (except for the same level shift) absent the epidemic and the shutdown.

Differences in average spending changes between Denmark and Sweden may be driven by differences in the socio-demographic composition of the sample of bank customers across the two countries.  To remove the influence of such differences and isolate the effect of the variation in lockdown policies, we estimate the following regression model:

\begin{equation}
\Delta spending_{ic}=\alpha_{c}+\mathbf{X}_{i}\delta_{c}+\epsilon_{ic}\label{eq:reg1}
\end{equation}

where $\alpha_{c}$ and $\mathbf{\delta}_{c}$ are country-specific parameter vectors and $\mathbf{X}_{i}$ is a vector of covariates capturing the age, sex, and permanent income of individual $i'$. Each variable is represented discretely by a set of one or more dummies. The model is fully saturated in the sense that we include all interactions between these dummies, as well as allowing for different coefficients in Denmark and Sweden.

We use the estimates from equation (\ref{eq:reg1}) to produce adjusted measures of the aggregate spending drop in each country: for each value of $c$, we compute predicted values from the model and take the average over individuals in the \textit{Danish} sample. For Denmark, the adjusted measure is just equal to the unadjusted $\Delta spending_{c}$. For Sweden, it is essentially a weighted average of individual spending changes where the weights are constructed to make the Swedish sample match the joint distribution of the variables in $\mathbf{X}_{i}$ in the Danish sample. One can think of this as an estimate of the counterfactual aggregate spending drop in a Denmark if it had followed the Swedish light-touch approach, holding socio-demograhic variables constant.

In order to estimate how the shutdown affects people differently depending on their health risk, we also construct measures of the drop in spending within age groups in each country. These measures are constructed exactly like the country-wide measures, except that excess spending for an individual in age group $a$ in country $c$ is now scaled relative to average daily spending in 2019 for individuals in that age group and country. Letting $\Delta spending_{iac}$ denote the change in excess spending for person $i$ belonging to age group $a$ in country $c$, we follow the same steps as above and compute the unadjusted averages $\Delta spending_{c}$ by averaging over individuals in each subsample. Similarly, we compute adjusted measures by estimating the equation:

\begin{equation}
\Delta spending_{iac}=\alpha_{ac}+\mathbf{X}_{i}\delta_{ac}+\epsilon_{iac}\label{eq:reg2}
\end{equation}

where $\alpha_{ac}$ and $\mathbf{\mathbf{\delta}}_{ac}$ are age-group-country-specific parameters and $\mathbf{X}_{i}$ is the same vector of controls as in equation (\ref{eq:reg1}), except age. For each age group, we then compute predicted values for both values of $c$ and evaluate them for the average individual within that age group in the Danish subsample. As for the full sample, the adjusted measures are equal to the unadjusted ones for Denmark, whereas the adjusted measures for Sweden can be thought of as providing age-specific counterfactual estimates for Denmark, holding the demographic composition of each age group fixed.


\section{Results}
\label{sec:results}

Figure 2 illustrates the high-frequency dynamics in aggregate spending in Denmark (upper panel) and Sweden (lower panel) and how it is affected by the pandemic. For each day of the window of analysis, we plot aggregate spending (red line) as well as aggregate spending on the reference day one year earlier (gray line), both scaled by average daily spending in 2019. In both countries, there is a pronounced weekly cycle with spending spikes around weekends and in Denmark there is moreover a pay day cycle with large spikes around the end of the month where most employees receive their salaries. In the beginning of the window, before the pandemic reached Scandinavia, both the level and the cyclicality of spending are strikingly similar to the reference period in both countries. In Denmark, spending drops sharply around the shutdown on 11 March 2020 and remains below the level in the reference period, as indicated by the shaded area, throughout the window of analysis. In Sweden, spending drops sharply at almost the exact same time although no significant restrictions were imposed. Presumably, this is no coincidence but reflects that the Danish shutdown responded to an escalating pandemic, similar in the two countries, with its own strong effect on spending. This highlights the empirical difficulty of separating the effects of social distancing laws and the pandemic they are designed to contain. 

Figure 3 shows our estimates of the drop in aggregate spending in each of the two countries relative to a counterfactual without the pandemic (red bars) and the differential drop in Denmark (blue bar). The estimated drop in Denmark is 28.74\% reflecting that average spending was 2.86\% \textit{above} the reference period before the shutdown and 25.88\% \textit{below} the reference period after the shutdown. The analogous estimate for Sweden is 24.84\% reflecting that average spending was 1.30\% \textit{above} the reference period before the shutdown and 23.54\% \textit{below} the reference period after the shutdown.\footnote{These estimates are reported with standard errors in Figure A1 in the Appendix. The estimates are based on weighting of the Swedish observations to match the socio-demographic (age, gender, income) composition of the Danish sample. The unweighted estimate of the spending drop in Sweden is 24.72\%.} The differential spending drop in Denmark is thus 3.9 percentage points and precisely estimated (the standard error is around 0.40 percentage points). This represents our estimate of the effect of social distancing laws on aggregate spending.

Making inference about the effect of social distancing laws based on a simple comparison of Denmark and Sweden raises a number of concerns. First, while we documented above that the two countries were similarly exposed to the spreading of the virus, one may be concerned that the ensuing global economic crisis affected them differently so as to confound our analysis. For instance, one may imagine that Swedish firms were particularly affected by the contraction of international trade and that differential stock market losses and unemployment risks can explain why Swedish consumers reduced spending so much despite the virtual absence of restrictions on economic activity. However, as shown in the Appendix, the major stock market indexes in Denmark and Sweden followed almost the exact same trajectory through the crisis (Figure A2) and, if anything, the rise in unemployment claims was slightly sharper in Denmark than in Sweden (Figure A3).\footnote{In a recent paper, we document that economic exposure to the COVID-19 crisis in the form of stock market participation and employment in at-risk industries only accounts for a limited part of the aggregate spending drop in Denmark (Andersen et al, 2020a).} Second, one may be concerned that the Danish government offered particularly generous subsidies to firms and workers and that these measures partly offset the negative effect of the social distancing laws in Denmark. In the Appendix, we compare the government programs designed to sustain the economy in Denmark and Sweden and show that they were strikingly similar. Finally, as there is some economic integration between Southern Sweden and Eastern Denmark, part of the drop in Swedish spending could, in principle, reflect that the Danish shutdown deterred Swedish consumers from spending in Denmark. However, the Swedes in our sample are not concentrated in Southern Sweden where access to Denmark is relatively easy, but spread out across the vast country. Before the pandemic, less than 1\% of card spending by the Swedish individuals in our sample was through points of payment in Denmark. 

Figure 4 illustrates how the effect of the shutdown, the differential spending drop in Denmark relative to Sweden, varies by age group.\footnote{The absolute magnitude of the effect in each country is reported with standard errors in Figure A4 in the Appendix}. The effect is negative for the youngest group (age 18-29): spending by this age group dropped around 10 percentage points more in Denmark than in Sweden. By contrast, the effect of the shutdown is positive for the oldest group (age 70+): spending by this age group dropped around 5 percentage points less in Denmark than in Sweden. In the intermediate age groups, the effect of the shutdown is generally moderately negative, sometimes statistically indistinguishable from zero. 

As the risk of suffering serious health consequences of the virus is strongly increasing in age, we interpret the age gradient as the manifestation of an underlying health risk gradient in the spending effects. Under this interpretation, the social distancing laws put severe constraints on the spending choices of individuals with low health risk: this group would have spent significantly more during the pandemic if the availability of goods and services with high social proximity had not been restricted despite the higher prevalence of the virus. At the same time, the social distancing laws created a safer environment for consumption by individuals with high health risk: this group would have spent significantly less during the pandemic if the shutdown had not contained the virus despite the higher availability of goods and services. 

Figure 5 shows the age gradient in the effect of the shutdown for categories of spending involving high or moderate social proximity: retail (i.e., high-street shops and malls), social spending (i.e., food, drink and entertainment away from home), personal care (e.g., hairdressers and dentists) and public transport (i.e., trains and busses).\footnote{The absolute magnitude of the effect in each country and each category is reported with standard errors in Figure A5 in the Appendix} Across all four categories, we observe a clear age gradient: the effect of the shutdown is strongly negative for the young (low health risk) and less negative, sometimes even positive, for the old (high health risk). 

These results provide insights about the intriguing positive effect of the shutdown on the total spending of groups with high health risk. They show that the overall positive effect reflects modest negative effects in categories with high social proximity where the shutdown severely limited availability (i.e., personal care and social spending) and positive effects in categories with  moderate social proximity where the shutdown imposed smaller constraints on availability (i.e., retail shopping and public transport). This illustrates, in a very concrete way, the mechanism through which social distancing can improve aggregate economic outcomes in a pandemic: by reducing the prevalence of the virus, it lowers the risk of contracting it in public transit and in retail shops, which stimulates the spending of individuals with a high underlying health risk.


\section{Conclusion}
\label{sec:conclusion}

This paper uses transaction-level bank account data from a large Scandinavian bank to study the effect of government social distancing laws on consumer spending in the COVID-19 crisis. We exploit a natural experiment to identify the effect of government restrictions separately from the impact of the virus and the health risks it entails: despite similar early experiences with the virus, the Danish government mandated social distancing to slow the spread of the disease while the Swedish government opted for a light-touch approach with a focus on voluntary recommendations. 

We present two key results. First, we estimate a massive reduction in consumer spending in both Denmark and Sweden around the date of the Danish lockdown. This drop consists of a common component to both countries of around 25 percent, and an additional drop in Denmark of just under 4 percentage points as a result of the shutdown. This finding suggests that the vast majority of the fall in economic activity in the COVID-19 crisis can be attributed to perceived disease risks influencing behaviour, rather than government restrictions. 

Second, we find that the shutdown in Denmark decreases spending among the young (aged under 29) by around 10 percent and \textit{increases} the spending of the old (aged over 70) by almost 5 percent, relative to a counterfactual without restrictions and with higher disease risks. This finding suggests that lockdowns have the potential to improve aggregate economic outcomes by constraining the choices of low-risk individuals (the young) who would otherwise contribute most to the spread of the disease and thus inducing high-risk individuals (the old) to choose a higher level of economic activity and mobility. 

Our analysis period ends on 5 April 2020. At the time of writing, excess deaths in Denmark have fallen to almost zero and the Danish government has eased social distancing laws. In contrast, Sweden has continued to experience excess mortality. Our results can only shine a light on the immediate effect of the COVID-19 crisis and the shutdown on economic activity. In future work, it will be interesting to to trace out the longer-run impact of the pandemic, initial shutdown policies, and subsequent re-opening.

\pagebreak


\thispagestyle{empty}\baselineskip.78\baselineskip


\clearpage

\begin{figure}
              \centering
                             \captionsetup{labelformat=empty}
                             \caption{\textbf{Table 1: Summary statistics.} \footnotesize This table presents summary statistics for our analysis sample of Danske Bank customers in Denmark (Column (1)) and Sweden (Column (3)) and the approximate population of Denmark (Column (2)) and Sweden (Column (4)) from which they are drawn. Statistics in Columns (1) and (2) are calculated in 2019. Population figures are sourced from Statistics Denmark (DST) and Statistics Sweden (SCB) for the most comparable population available on their online statistics banks: 18+ year olds in 2018. Some differences in variable construction are explained below. \\
*For the Danish sample, disposable income is calculated based on a classification of account inflows and is the sum of inflows labelled salary, capital income, government transfers, and pension. For the Swedish sample, income is imputed based on average spending in 2019, assuming that the Swedish sample has the same \textit{average propensity to consume} as the Danish sample.
**Individual-level averages of the 20+ population in 2018.
Total spending is measured as the sum of debit and credit card transactions, mobile wallet payments, cash withdrawals, and electronic invoices associated with online shopping. Details on the construction of the spending categories can be found in the Appendix.}
                             \label{Fig:DiscByYear}
                             \centering
                   \includegraphics[width=.95\textwidth]{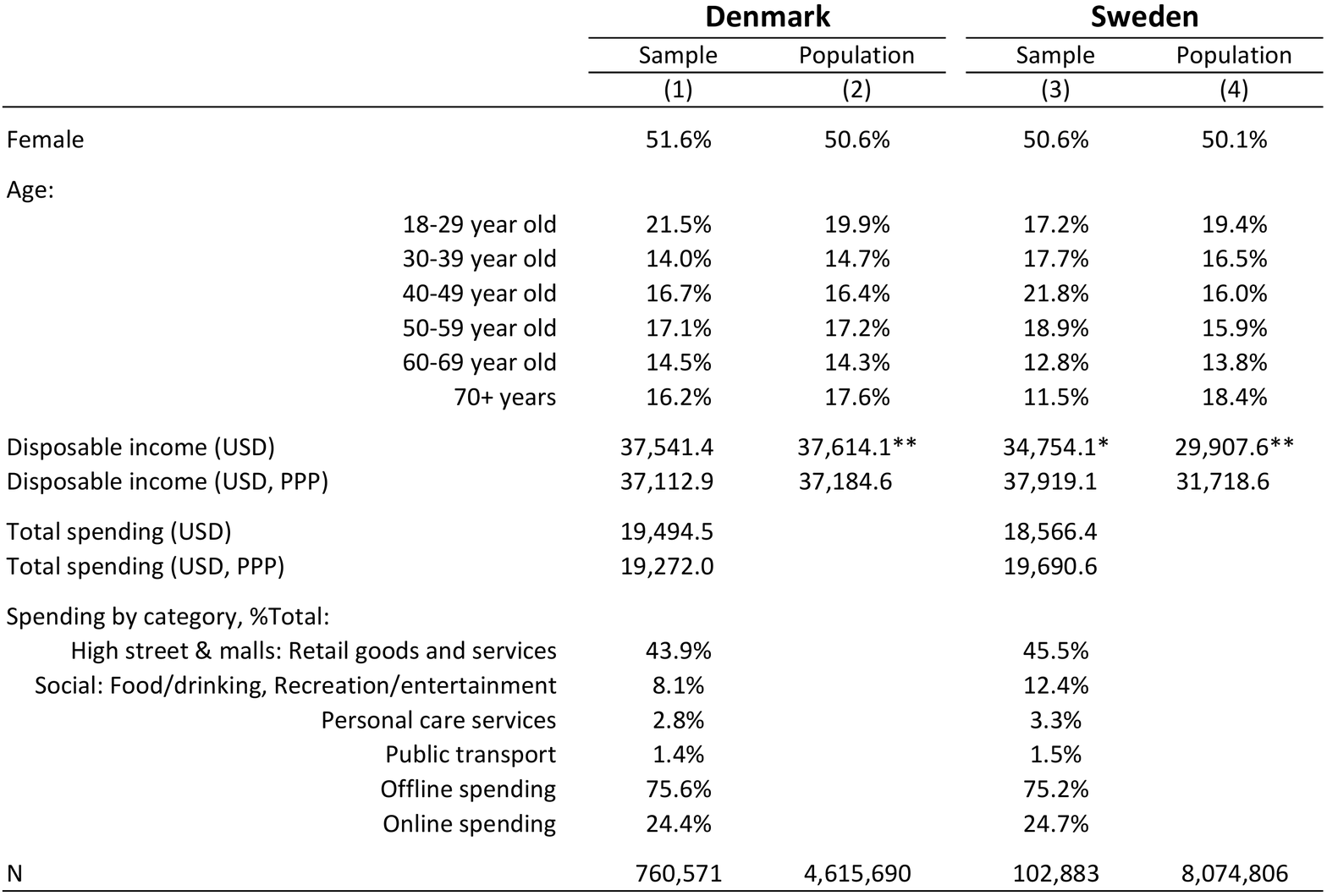}
\end{figure}

\begin{figure}
              \centering
                             \captionsetup{labelformat=empty}
                             \caption{\textbf{Figure 1: Impact of the Covid-19 outbreak on mortality and health.} \footnotesize The figure shows weekly excess mortality in Denmark, Sweden and Italy for weeks 2 through 15 in 2020 (top panel) and excess Google searches for symptoms of illness (bottom panel) for the same period. Excess mortality for each country is calculated based on daily data on deaths sourced from Statistics Denmark (DST, Denmark), Statistics Sweden (SCB, Sweden) and ISTAT (Italy). Deaths data for Denmark and Sweden cover the whole population, whereas the Italian data is available for approximately half of all Communes. Excess mortality is calculated as the percentage difference between the total number of deaths from all causes in a week in 2020 and the average number of total deaths in that week, 2015-2019. Excess Google searches are calculated based on data from Google Trends of weekly indices of search intensity for the terms ``cough" (DK: ``hoste", SE: ``hosta", IT: ``tosse"), ``fever" (``feber", ``feber", ``febbre"), ``sick"  (``syg", ``sjuk", ``malato"). Individual indices for each search term are summed in order to create a composite index of all terms. Excess Google searches are then calculated in the same way as for excess mortality.}
                             \label{Fig:DiscByYear}
              \includegraphics[width=0.9\textwidth]{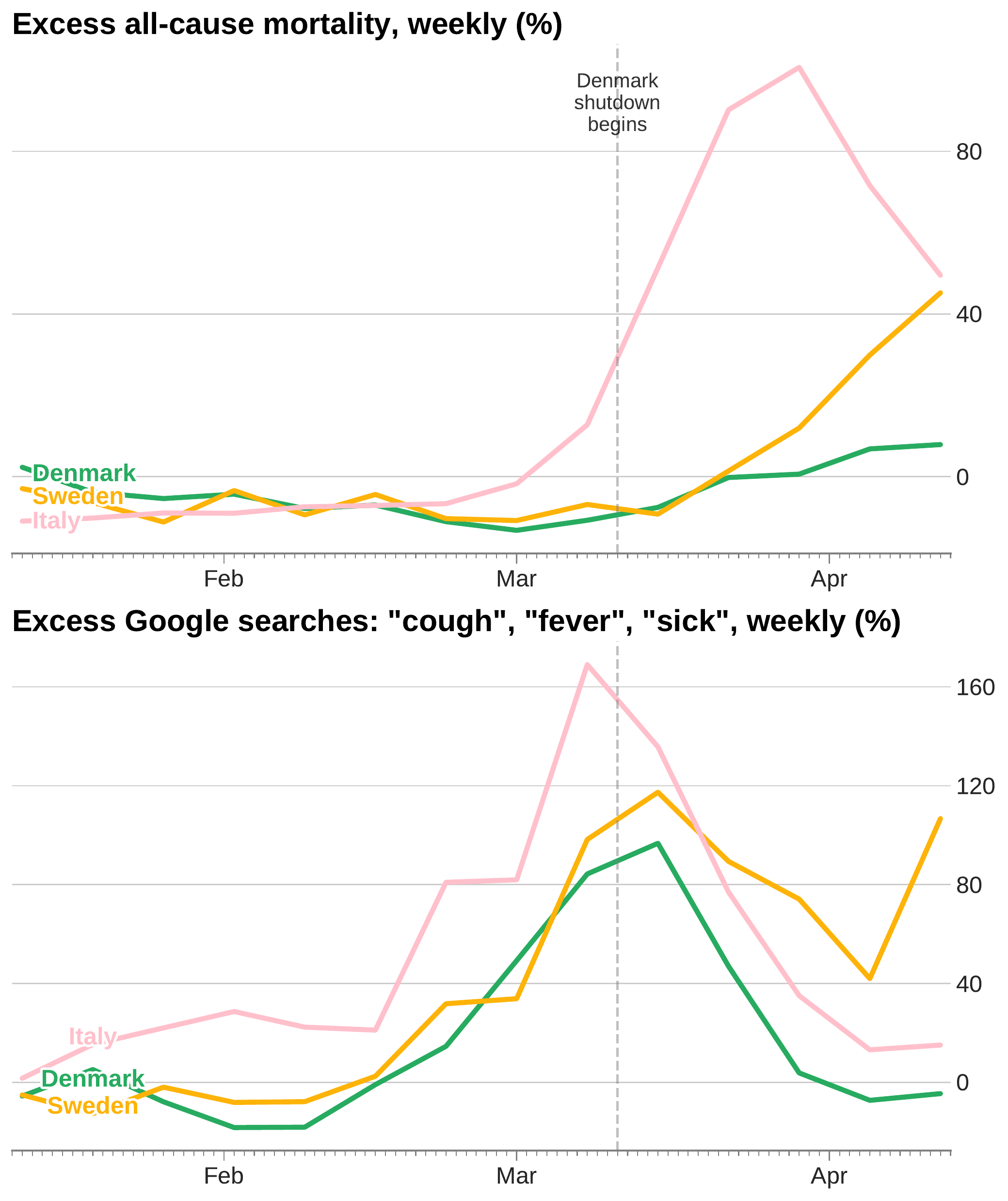}
\end{figure}

\pagebreak

\begin{figure}
              \centering
                             \captionsetup{labelformat=empty}
                             \caption{\textbf{Figure 2: Evolution of aggregate spending in Denmark and Sweden.} \footnotesize The figure shows the evolution of aggregate spending in Denmark and Sweden from 2 January to 5 April 2020. Red lines show the evolution of spending in 2020 as a percentage of daily average spending in 2019. Grey lines show the same series for the same weekday in 2019, i.e., 364 days earlier. The dash vertical line denotes 11 March, when the Danish government announced the lockdown. Shaded red regions highlight the drop in spending in both countries at this point in time.}
                             \label{Fig:DiscByYear}
              \includegraphics[width=0.9\textwidth]{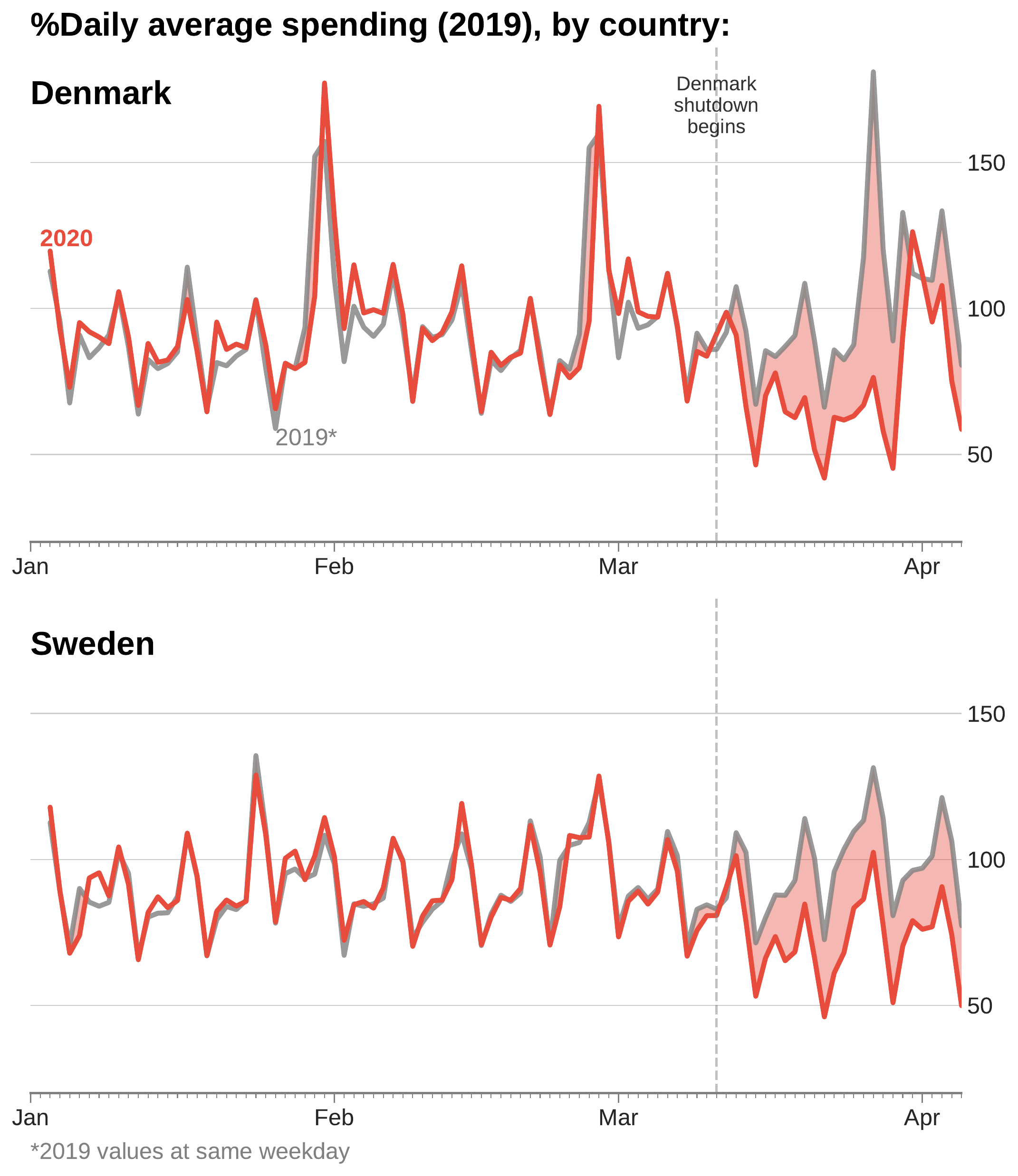}
\end{figure}

\pagebreak

\begin{figure}
              \centering
                             \captionsetup{labelformat=empty}
                             \caption{\textbf{Figure 3: Effect of the shutdown on total spending.} \footnotesize The figure shows the impact of the COVID-19 crisis on consumer spending in Denmark (DEN) and Sweden (SWE) (top panel) and isolates the effect of the Danish shutdown (bottom panel). The estimates are based on weighting of the Swedish observations to match the socio-demographic (age, gender, income) composition of the Danish sample, as described in our empirical framework. Weights are based on 6 age groups (ages 18-29, 30-39, 40-49, 50-59, 60-69, and 70+), sex, and quartiles of purchasing power parity (PPP) adjust permanent income. Confidence bounds at the 95\% level (black vertical lines) are based on robust standard errors.}
                             \label{Fig:DiscByYear}
              \includegraphics[width=.9\textwidth]{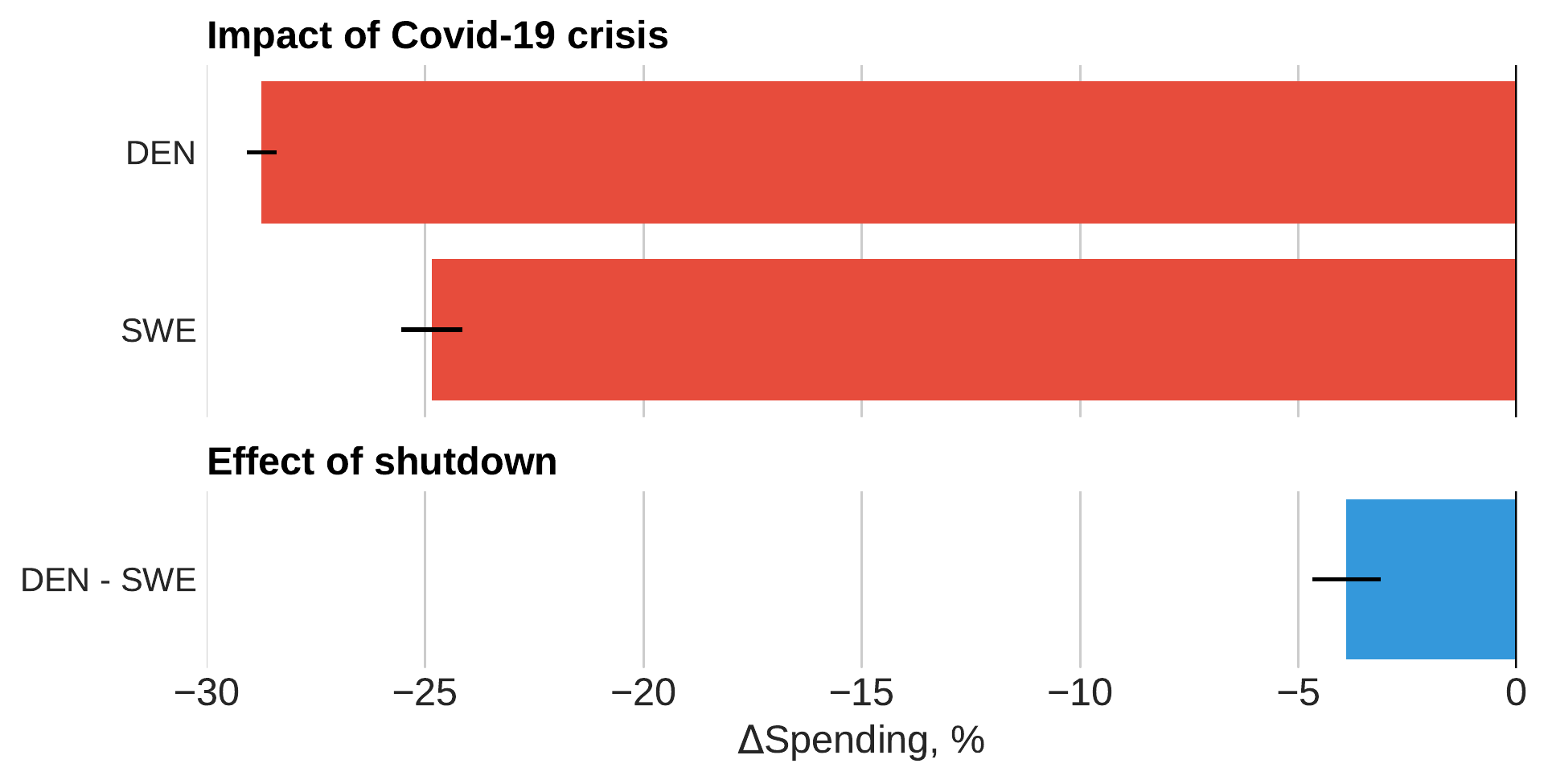}
\end{figure}

\pagebreak

\begin{figure}
              \centering
                             \captionsetup{labelformat=empty}
                             \caption{\textbf{Figure 4: Effect of the shutdown by age group.} \footnotesize The figure shows the effect of the shutdown on total spending by age group, a measure of COVID-19 disease risk. Age-specific estimates of the shutdown effect are calculated as the difference between the estimated drop in consumer spending among that age group in Denmark and Sweden, with the Swedish sample weighted to match the socio-demographic composition of the Danish sample. Weights are based on 6 age groups (ages 18-29, 30-39, 40-49, 50-59, 60-69, and 70+), sex, and quartiles of purchasing power parity (PPP) adjust permanent income. Confidence bounds at the 95\% level (black vertical lines) are based on robust standard errors. Estimates of the country-specific effects on total spending for each age group are provided in Figure A4 in the Appendix.}
                             \label{Fig:DiscByYear}
              \includegraphics[width=1\textwidth]{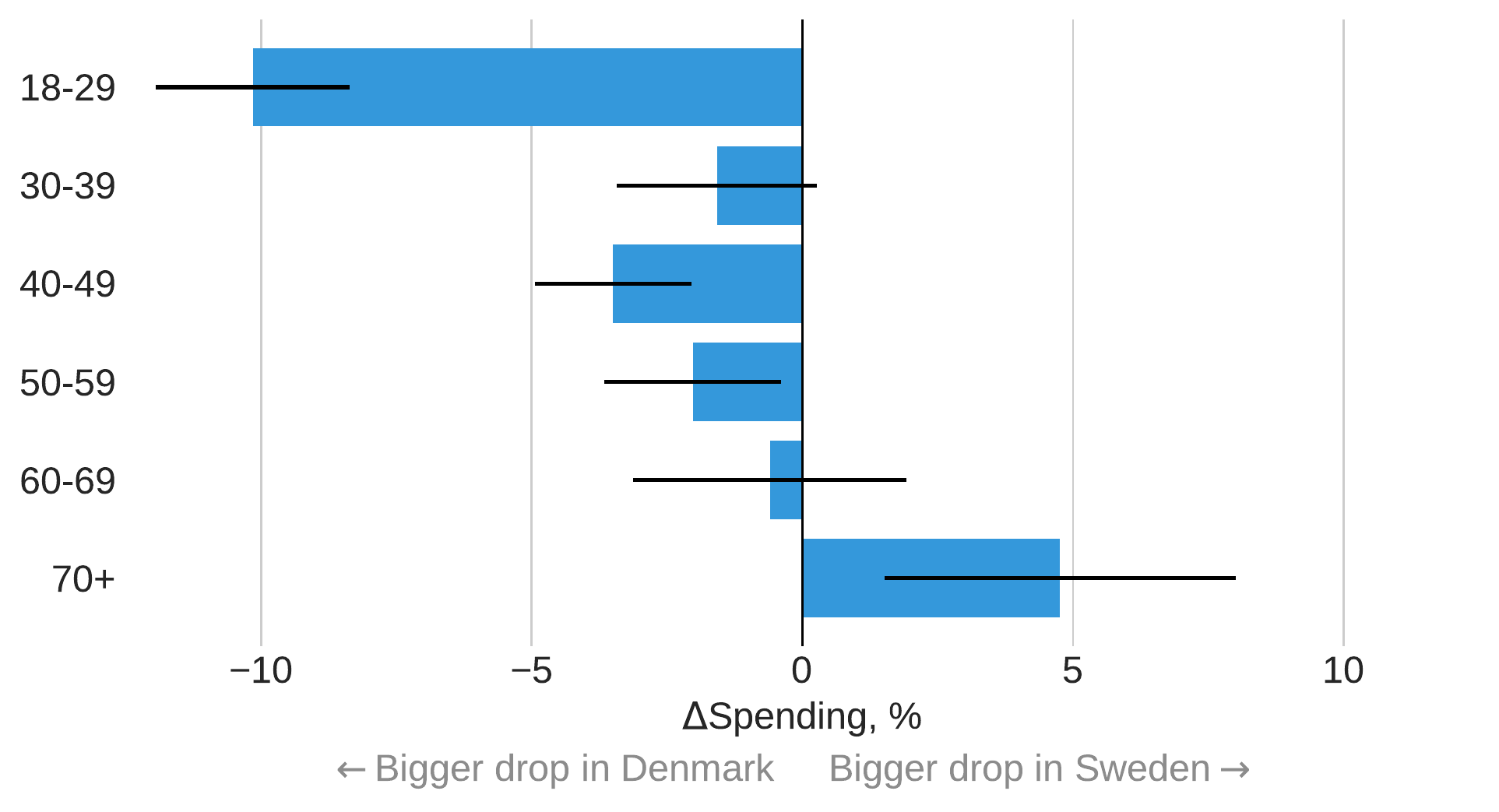}
\end{figure}

\pagebreak

\begin{figure}
              \centering
                             \captionsetup{labelformat=empty}
                             \caption{\textbf{Figure 5: Effect of the shutdown by age group on spending categories that vary by social proximity.} \footnotesize The figure shows the effect of the shutdown by age group and across categories of offline spending that vary in their degree of social proximity. Confidence bounds at the 95\% level (black vertical lines) are based on robust standard errors. Details on the spending categories are provided in Table A1 in the Appendix. Estimates of the country-specific effects on spending categories for each age group are provided in Figure A5 in the Appendix.}
                             \label{Fig:DiscByYear}
              \includegraphics[width=1\textwidth]{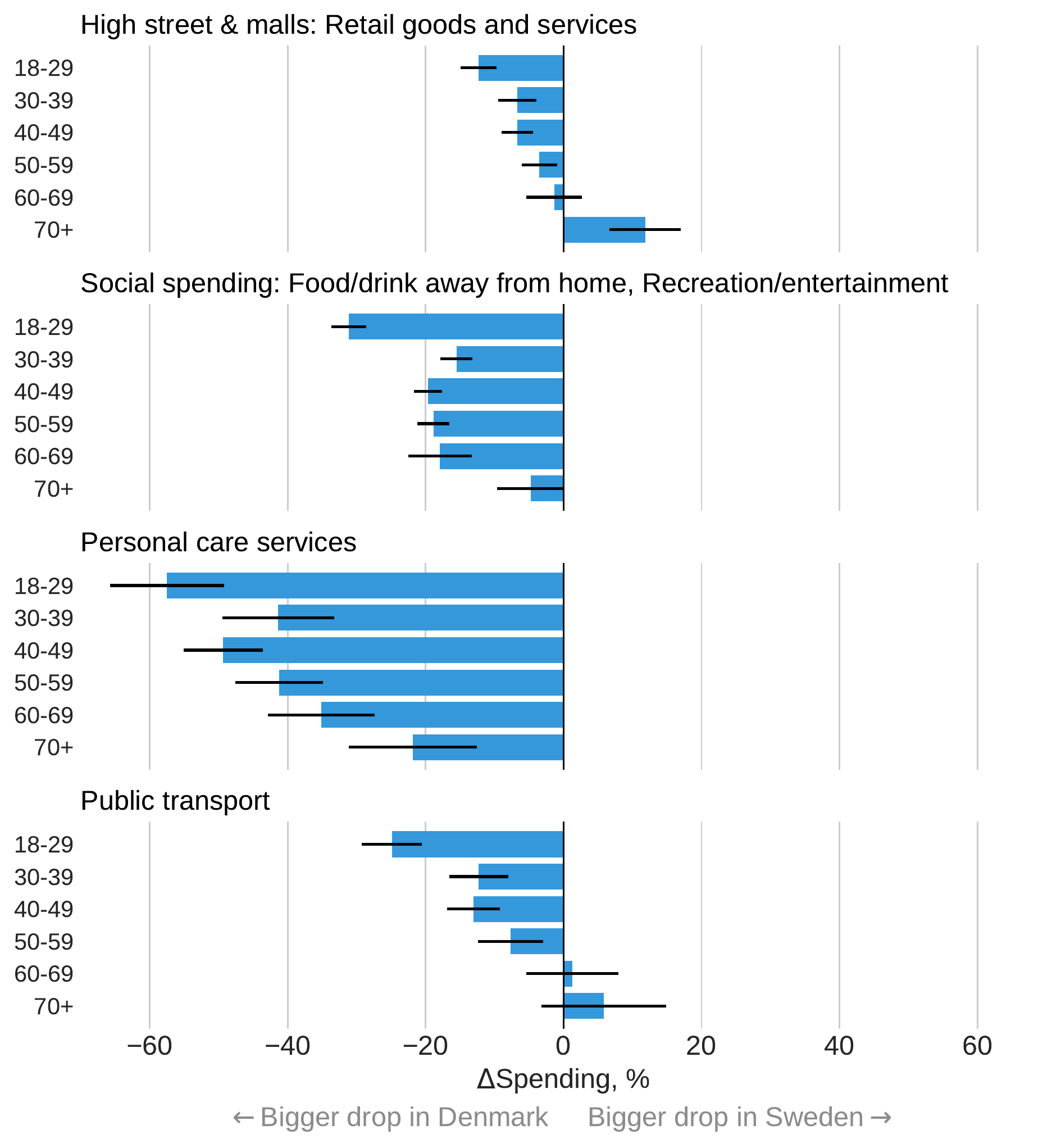}
\end{figure}

\pagebreak


\clearpage
\vspace*{\fill}
\begin{center}
              \begin{minipage}{.8\textwidth}
                             \centering{{\huge ONLINE \vskip 5mm APPENDIX}}
              \end{minipage}
\end{center}
\vfill 
\pagebreak
\appendix

\begin{figure}
              \centering
                             \captionsetup{labelformat=empty}
                             \caption{\textbf{Table A1: Construction of spending categories} \footnotesize This table shows the grouping of Merchant Category Codes (MCCs) into spending categories. We only report MCCs accounting for more than 0.5\% of all spending within each category. This categorisation applies to offline, i.e., in-merchant, payments only.} 
                             \label{Fig:DiscByYear}
                \includegraphics[width=.96\textwidth]{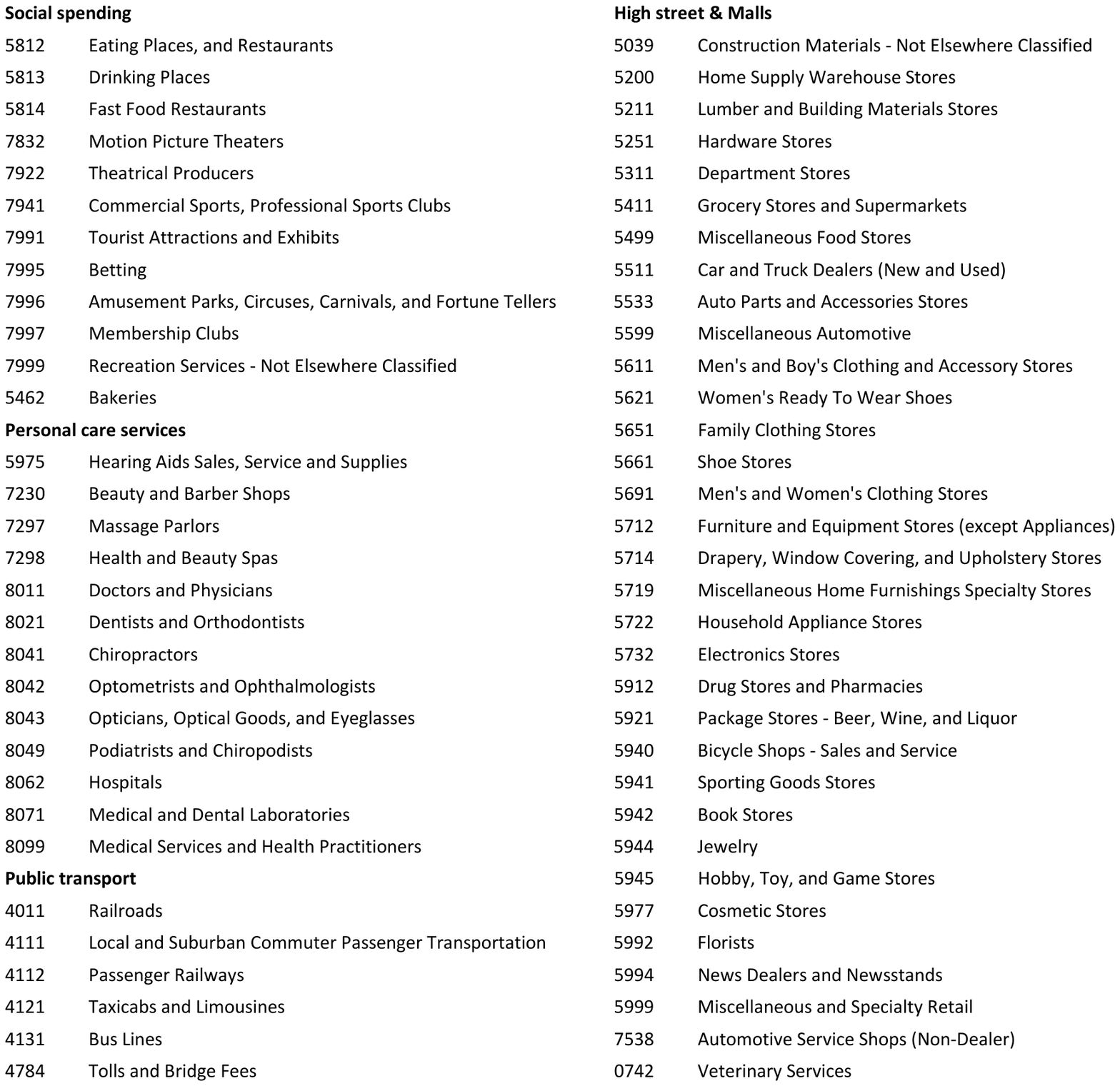}
\end{figure}

\pagebreak

\begin{figure}
              \centering{
                             \captionsetup{labelformat=empty}
                             \caption{\textbf{Figure A1: Effect of the shutdown in Denmark and Sweden} \footnotesize This figure breaks down our estimate of the effect of social distancing laws in Figure 2: gray bars denoted \textit{Pre} show excess spending in the period 2 January - 15 February; gray bars denoted \textit{Post} show excess spending in the period 12 March - 5 April; and red bars denoted \textit{Post-Pre} show the difference between them and thus represent our estimates of the total spending drop induced by the COVID-19 crisis in Denmark (DEN) and Sweden (SWE) respectively. Finally, the blue bar denoted \textit{Shutdown effect} shows the difference between the red bars and thus represent our estimate of the total spending drop induced by the Danish shutdown. The estimates are based on weighting of the Swedish observations to match the socio-demographic (age, gender, income) composition of the Danish sample, as described in our empirical framework. Weights are based on 6 age groups (ages 18-29, 30-39, 40-49, 50-59, 60-69, and 70+), sex, and quartiles of purchasing power parity (PPP) adjust permanent income. Confidence bounds at the 95\% level (black vertical lines) are based on robust standard errors.                   
                             } 
                             \label{Fig:DiscByYear}
              \includegraphics[width=1\textwidth]{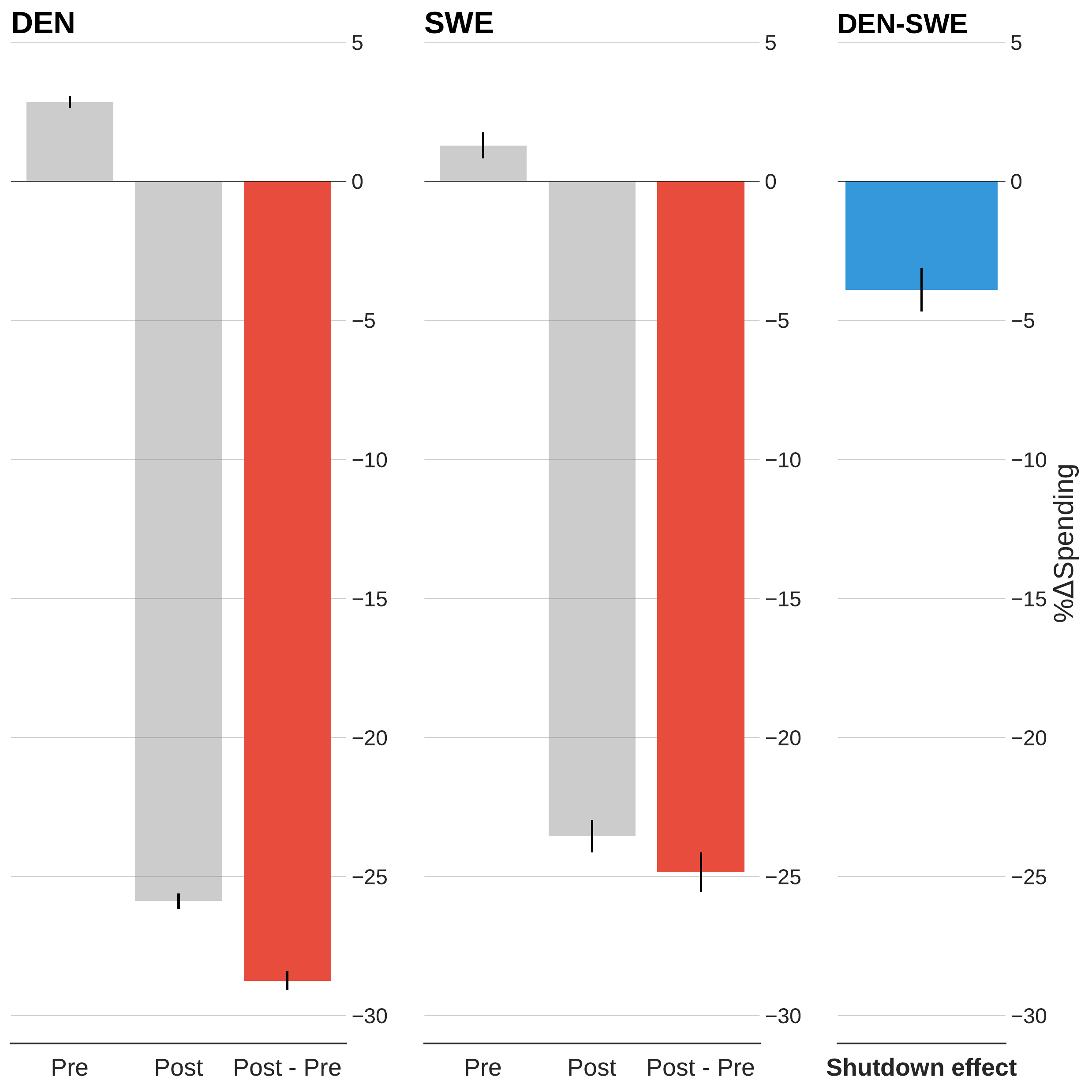}}
\end{figure}

\begin{figure}
              \centering
                             \captionsetup{labelformat=empty}
                             \caption{\textbf{Figure A2: Effect of the Covid-19 crisis on stock markets} \footnotesize This figure shows the effect of the Covid-19 crisis on the Danish (OMXC25) and Swedish (OMXS30) stock market indices. Source: Nasdaqomxnordic.com.} 
                             \label{Fig:DiscByYear}
              \includegraphics[width=.9\textwidth]{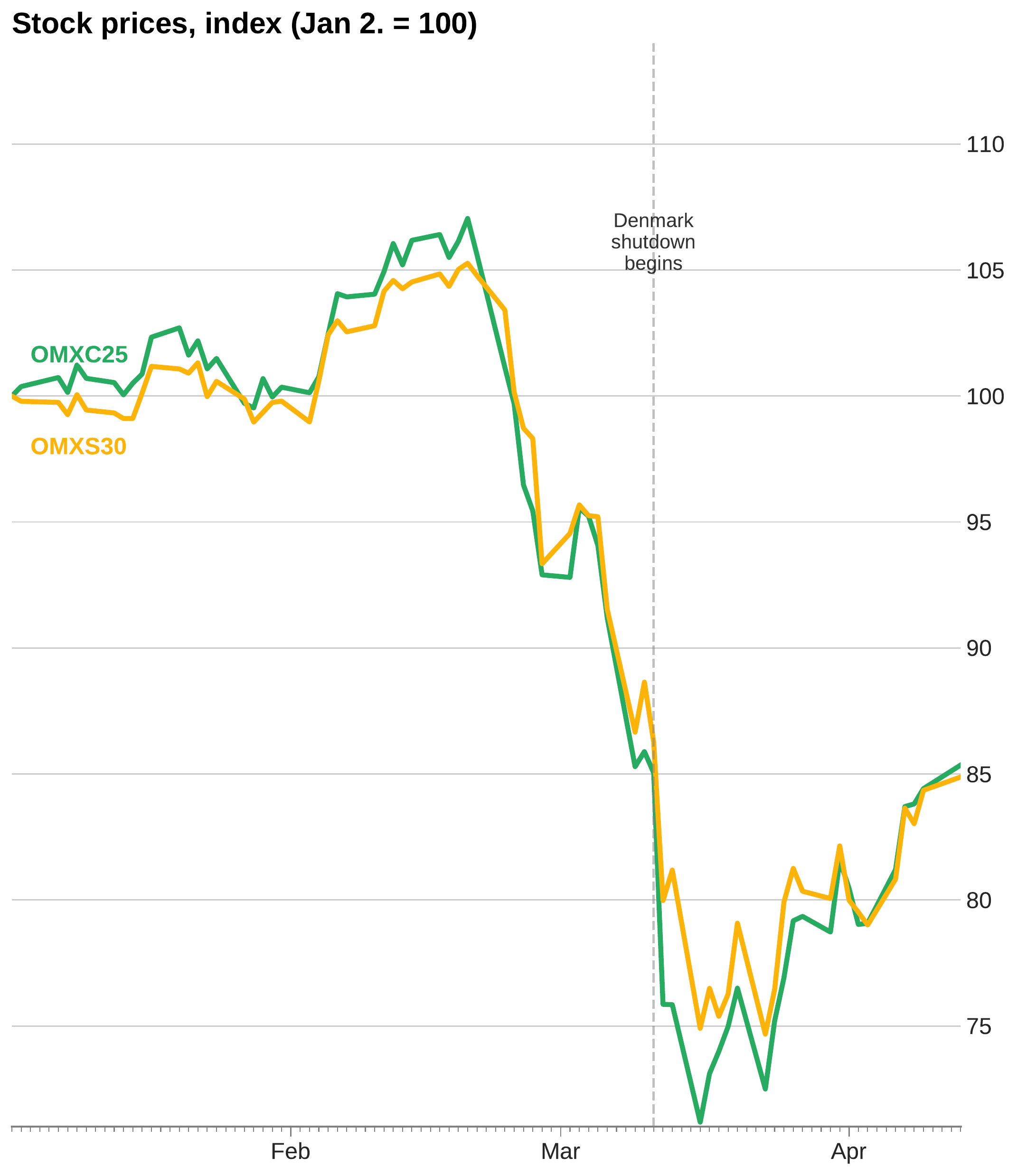}
\end{figure}

\pagebreak

\begin{figure}
              \centering
                             \captionsetup{labelformat=empty}
                             \caption{\textbf{Figure A3: Effect of the Covid-19 crisis on new unemployment claims} \footnotesize This figure shows the effect of the Covid-19 crisis on new unemployment claims in Denmark and Sweden. The series show the percentage deviation in weekly new unemployment claims in 2020 from the historical average in the same week over 2015-2019 for Denmark and 2018-2019 for Sweden. Source: Statistics Denmark and Statistics Sweden.} 
                             \label{Fig:DiscByYear}
              \includegraphics[width=.9\textwidth]{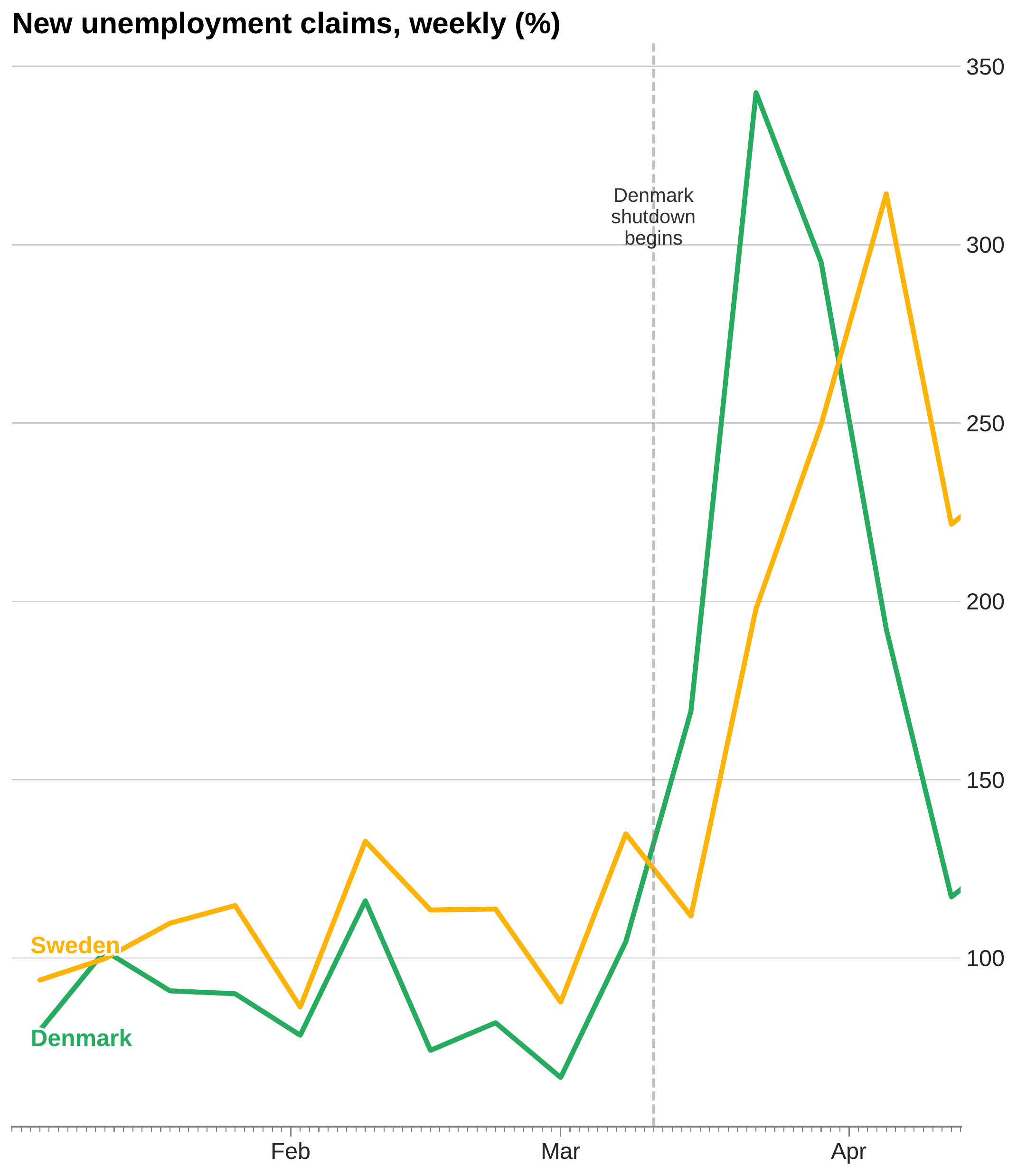}
\end{figure}

\pagebreak
\clearpage
\begin{figure}
              \centering{
                             \captionsetup{labelformat=empty}
                             \caption{\textbf{Figure A4: Effect of the shutdown by age group} \footnotesize This figure shows the country-specific estimates underlying Figure 4.} 
                             \label{Fig:DiscByYear}
              \includegraphics[width=1\textwidth]{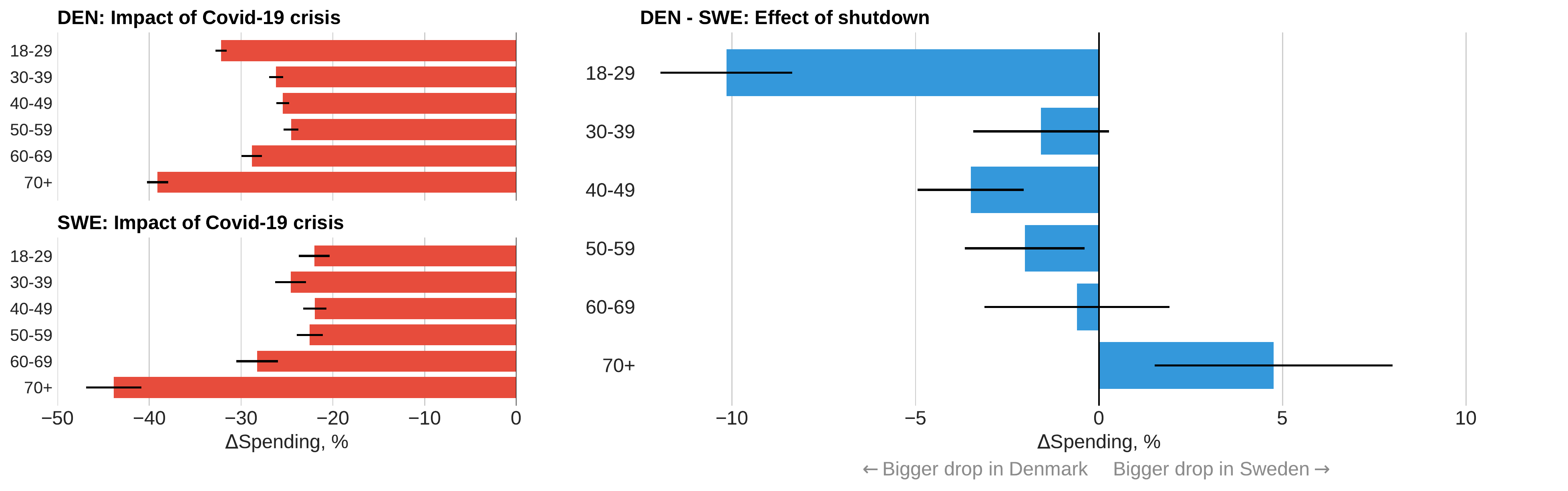}}
\end{figure}

\pagebreak
\clearpage

\begin{figure}
              \centering
                             \captionsetup{labelformat=empty}
                             \caption{\textbf{Figure A5: Effect of the Covid-19 crisis on spending categories} \footnotesize This figure shows the country-specific estimates underlying Figure 5.} 
                             \label{Fig:DiscByYear}
              \includegraphics[width=.62\textwidth]{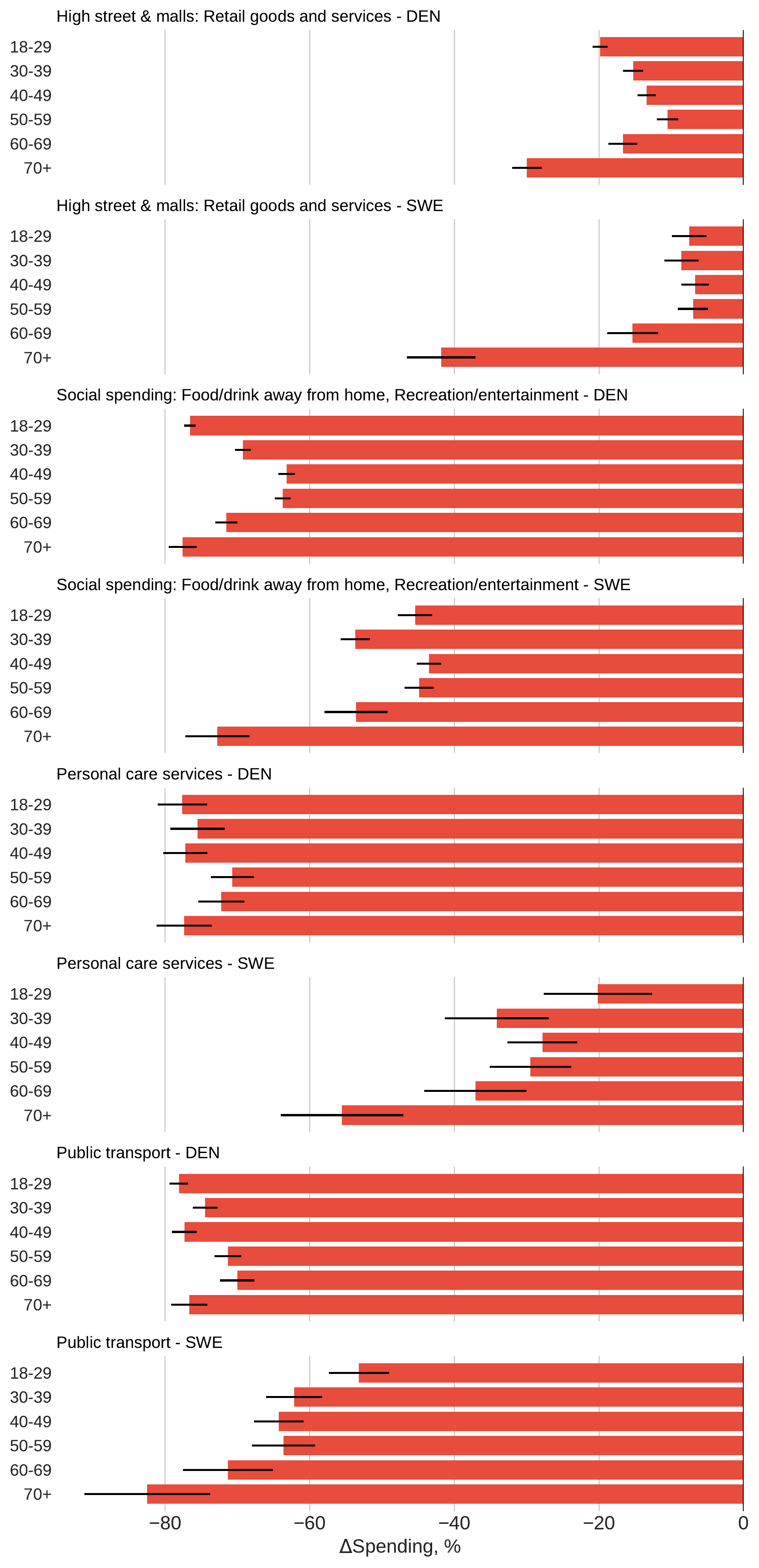}
\end{figure}

\pagebreak
\clearpage


\noindent\textbf{Economic policy responses}

Despite very different policy responses to managing the spread of Covid-19, Sweden and Denmark have introduced similarly massive government programs to mitigate the financial damage of the crisis to businesses and households. First, both countries have introduced significant loan subsidies, including state guarantees of 70\% of new corporate loans related to Covid-19. Second, extensive furlough support schemes were introduced in order to prevent mass lay-offs. In Denmark, the government committed to pay 75\% of the salary of private sector employees who are sent home but kept on the payroll. In Sweden, the government guarantees all workers will receive up to 90\% of their salary while allowing employers to cut working hours by up to 80\%. Third, in both countries the government will provide substantial cost subsidies that cover up to 80\% of fixed costs in Denmark and 75\% in Sweden. Finally, both governments are allowing many companies to postpone VAT payments.

The fact that both countries have introduced such similar and extensive financial support to firms and households suggests that both governments expected the Covid-19 crisis to result in significant economic damage. We have provided direct evidence on this, showing how both countries experienced massive drops in consumer spending. Other evidence points to similar effects across countries in terms of stock market and labor market performance. Figure A3 shows that both countries experienced highly similar trends in stock market performance in January and early February 2020 and then experienced the same sudden drop in their stock markets. Similarly, Figure A4 shows that both countries experienced highly similar trends in new unemployment claims in January and February 2020 and then experienced the same sudden spike in claims in March.

\end{document}